# The Effect of Computer-Generated Descriptions on Photo-Sharing Experiences of People With Visual Impairments

YUHANG ZHAO, Information Science, Cornell Tech, Cornell University
SHAOMEI WU, Facebook Inc.
LINDSAY REYNOLDS, Facebook Inc.
SHIRI AZENKOT, Information Science, Cornell Tech, Cornell University

Like sighted people, visually impaired people want to share photographs on social networking services, but find it difficult to identify and select photos from their albums. We aimed to address this problem by incorporating state-of-the-art computer-generated descriptions into Facebook's photo-sharing feature. We interviewed 12 visually impaired participants to understand their photo-sharing experiences and designed a photo description feature for the Facebook mobile application. We evaluated this feature with six participants in a seven-day diary study. We found that participants used the descriptions to recall and organize their photos, but they hesitated to upload photos without a sighted person's input. In addition to basic information about photo content, participants wanted to know more details about salient objects and people, and whether the photos reflected their personal aesthetic. We discuss these findings from the lens of self-disclosure and self-presentation theories and propose new computer vision research directions that will better support visual content sharing by visually impaired people.

## 1 INTRODUCTION

Sharing memories and experiences via photos is a common way to engage with others on social networking services (SNSs) [39,46,51]. For instance, Facebook users uploaded more than 350 million photos a day [24] and Twitter, which initially supported only text in tweets, now has more than 28.4% of tweets containing images [39]. Visually impaired people (both blind and low vision) have a strong presence on SNS and are interested in sharing photos [50]. They take photos for the same reasons that sighted people do: sharing daily moments with their sighted friends and family [30,32]. A prior study showed that visually impaired people shared a relatively large number of photos on Facebook—only slightly less than their sighted counterparts [50].

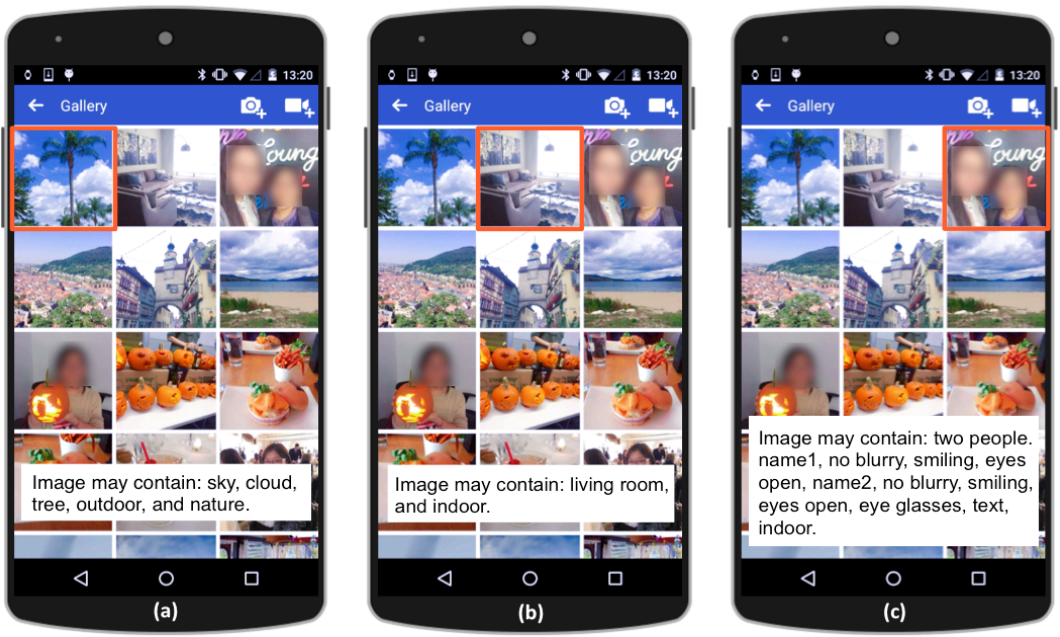

Fig. 1. Computer-generated descriptions in Facebook. The text in the white boxes is the descriptions that are read to a blind user by TalkBack. The descriptions are normally invisible. We show it visually to demonstrate the design.

While they want to share photos with others, visually impaired people find it difficult to understand the contents of photos and select good photos to post from their albums [30,47]. If a visually impaired person takes a photo and does not upload it immediately, it is hard for her to navigate through the album and find that photo independently, especially when many photos have accumulated in her album over time. Moreover, it is difficult to judge the quality of a photo, for example, to determine whether the photo is blurry, whether a person has her eyes closed in the photo, or whether the photo is aesthetically pleasing [30].

Researchers and designers have tackled this problem, proposing techniques to help people with visual impairments access and understand the contents of a photo. Most of these efforts used human-powered services (e.g., crowd workers, friends) to provide photo descriptions or answer a photo-based question [13,15,54,55]. However, such systems are hard to scale and sustain due to the limited number of volunteers, the monetary cost of crowd workers, and the possible social costs involved when asking friends [18,22]. Also, since there can be private information in local photos without the user knowing it, human-powered services also exposed visually impaired users to high privacy risks when sending personal photos to human assistants [16]. Therefore, other approaches are necessary for captioning a number of personal photos from a user's album in a photo-sharing use case.

Recent research by Wu et al. presented automatic alt-text (AAT) [51], which used computer vision to generate descriptions for photos on the Facebook Newsfeed. They showed that AAT helped visually impaired people to interpret photos on Facebook, making Facebook more engaging for them. MacLeod et al. [37] conducted user studies to understand how visually impaired people perceive computer-generated image descriptions on Twitter, showing that people tend to trust the descriptions even when the descriptions and the tweet text are incongruent. These projects investigated computer-generated image descriptions for *consuming* visual content on SNSs for people with visual impairments, but did not address visual content *production.* Sharing photos on SNSs is an act of self-disclosure with unique risks and rewards [3,10]. It is not yet known how computer-generated descriptions would impact photo sharing for visually impaired users. In this paper, we explore how visually impaired people use and experience computer-generated descriptions in a photo-sharing context. We also aim to understand what qualities define a "good" description for this particular use case.

We conducted an interview study with 12 visually impaired participants to understand their experiences with photo sharing on SNSs, including challenges, strategies, and specific photo-description preferences. Guided by our findings, we enhanced the photo-sharing feature in the Facebook mobile application by adding computer-generated descriptions to the photos in local albums to help visually impaired users post photos to Facebook. We developed a fully functional prototype of the feature, which is now available on the Facebook Android application1.

To understand the effect of the enhanced photo-sharing feature, we conducted a seven-day diary study with six visually impaired participants. Our study showed that computer-generated descriptions were useful in terms of helping people recall memories and organize local photos. However, our participants did not yet completely rely on the generated descriptions and still asked for a sighted person's input before uploading photos. We discuss their behaviors through the lens of self-disclosure and self-presentation theories. Our study further identified new challenges for current technologies to fully support their needs. Compared to the content consumption scenario, our participants had much higher demands for accuracy and the richness of computer-generated descriptions when curating and sharing photos. They wanted to know more details about salient objects and people (e.g., behaviors of people and animals, characteristics of objects), as well as whether the photo reflected their personal aesthetic. Our findings shed light on new directions for computer vision research to better support visual content sharing for visually impaired people, such as evaluating the artistic quality of a photo based on individual taste, and distinguishing subtle differences between similar photos.

In summary, this paper has three contributions:
1. The design and evaluation of an accessibility feature in the Facebook mobile application that provides computer-generated photo descriptions to visually impaired people when sharing photos.
2. Insights on how computer-generated descriptions affect visually impaired people's photo-sharing experiences, analyzed in the context of self-disclosure and self-presentation theories.
3. Directions for computer vision research to better support visual content sharing for visually impaired people.

## 2  RELATED WORK

People with visual impairments take and share photographs for the same reasons that sighted people do [30,32], however, they encounter more challenges than sighted people both in taking photos and browsing their albums to select good photos [46,50]. Researchers have developed several methods to assist blind photography, including audio feedback that facilitates aiming the camera [32,45], and recorded memos and ambient sounds that augment photos [2,30]. Unlike this prior work on enhancing the photo-taking process, we focus on generating descriptions to help visually impaired people better browse and select photos for sharing. Prior research on generating photo descriptions took one of two approaches: human computation [13,17,18,55] and computer vision [32,51,56]. We describe research following each approach below.

### 2.1  Human-Powered Photo Descriptions

Most prior research [13,17,18] and products [55] used human-powered services, such as friends or crowd workers, to describe a photo or answer questions about a photo for visually impaired users. For example, VizWiz [13] is a mobile inquiry application that allows a visually impaired user to upload photos with audio questions and receive answers from sighted workers or friends. However, someone must pay the crowd workers to sustain the application. Using social networks instead of anonymous crowdworkers, referred to as friendsourcing [12], is a low-cost way to get information about photos for visually impaired people. VizWiz Social [18] used friendsourcing by allowing a user to send photos and questions to her friends via email, Facebook, or Twitter. However, a user study showed that blind users did not prefer this method because of its high social costs that arose from burdening friends or not appearing independent.

---

[1] Unlike the prototype used in the study, the final version does not include face recognition. See Section 'Enhancing the Photo-Sharing Feature.'

To reduce the social and monetary cost, Brady et al. [17] introduced social microvolunteering, in which questions from blind people can be posted on the volunteers' Facebook newsfeed so that all the volunteers' Facebook friends can help answer the questions. This method could provide an answer in around six minutes if a question was posted to nine volunteers' newsfeeds. However, this system raised privacy concerns because some personal and private information on photos might be unintentionally revealed to human assistants.

Although human-powered systems can provide high quality photo captions, they are hard to speed up, scale and sustain [51] because of the limited number of volunteers, the financial cost, and the possible social cost. Moreover, having other people see one's local photos raises privacy concerns and exposes visually impaired users to higher privacy risks [16]. Thus, other options are needed to obtain information about local photos for visually impaired people.

## 2.2 Computer-Generated Photo Descriptions

Another way to provide photo descriptions for visually impaired people is to use computer vision technology. Some prior work focused on generating automatic captions for online photos to help visually impaired people consume photo-based information. For example, Alt Text Bot [57] is a browser extension that uses image recognition technology to provide descriptions for images on the web, which helps blind users better understand web content. Automatic alt-text [51] uses computer vision technology to detect faces and recognize objects and themes in photos on Facebook. MacLeod et al. [37] explored visually impaired people's experience with computer-generated descriptions on Twitter and suggested that descriptions should be framed in a way that would encourage users' skepticism when recognition results were uncertain. While these research efforts explored how people consume visual content online, it is still unknown how such technology affects visually impaired people's experience in producing visual content (e.g., sharing photos).

Some existing applications used computer vision technology to describe users' local photos, but no formal studies were conducted to understand how these descriptions affected photo-sharing behaviors. Google Goggles [56] is a mobile application that provides the user information about a photo, such as whether it contains a famous landmark. However, this tool serves as an independent service and requires a user to explicitly upload each photo from an album, which may not fit for investigating a number of photos in a photo-sharing scenario. Moreover, there is no known formal evaluation of its effectiveness. Another example is EasySnap [32,49], an iPhone application that assists blind photography by indicating the status of a current frame with audio feedback (e.g., the position of an object). It has an accessible album, which extracts location information from Google Maps and recognizes the photo content with AI technology to generate a photo description at the time a photo is taken. Jayant et al. [32] evaluated the effectiveness of EasySnap for taking good photos, but did not explore its effect on photo sharing.

We seek to understand how computer-generated photo descriptions affect visually impaired people's experience in uploading photos to SNSs, including whether the computer-generated descriptions promote photo-based information production, the impact of recognition failures on users' experience, and their needs (e.g., accuracy, details) for caption quality in a photo sharing scenario. To our knowledge, no prior work has focused on this before.

## 2.3 Self-Disclosure and Self-Presentation On SNSs

We use self-presentation and self-disclosure theories as a framework to understand the photo-sharing behaviors of visually impaired people and their experience with computer-generated descriptions.

Self-presentation, also known as impression management, refers to a person's strategic behaviors to "convey an impression to others which is in his interests to convey" [26]. It often involves keeping one's "true" self private, while exaggerating one's own perceived favorable attributes or behaviors [7]. Two main reasons motivate self-presentation behaviors: pleasing one's audience and self-construction. Pleasing one's audience involves matching one's self-presentation to the audience's expectations and preferences of that individual, whereas self-construction involves matching one's self-presentation to one's own ideal self [6,8]. These motivations imply the potential rewards from self-presentation, such as building relationships with others, or gaining benefits (e.g., power, respect, affinity) from a desirable public image [33].

As SNSs have become a major platform for social interaction [5], researchers have been using self-presentation theory to explain people's behaviors on SNSs, like Facebook. Self-presentation is one major motivation for Facebook use [40]. People engage in self-presentation in different ways, when sharing photos, filling in profile information, and composing wall posts [53]. Although profile information reflects users' self-image most [4,14], sharing photos can also present oneself in certain ways, constructing an identity, changing people's perspectives, and even making one's self-image iconic [21].

Self-disclosure is closely related to self-presentation. It is defined as the act of "revealing intimate information about oneself to others" [28]. It is involved in some self-presentation strategies (e.g., self-promotion) [36] and can help people achieve the rewards of self-presentation. However, it also carries potential risks, such as vulnerability and information loss [3].

People engage in self-disclosure to mass audiences on SNSs through non-direct status updates (e.g., posts on the Facebook Newsfeed) [9]. They derive benefits from such activities that include social validation, relational development, and self-expression [10]. Meanwhile, the large and ill-defined audience on SNSs also poses challenges, such as loss of privacy, increased vulnerability [10,29], and unintended presentations of self [23]. To better explain the dynamics of information disclosure on SNSs, Bazarova and Choi [10] introduced the functional model of self-disclosure, which takes into account both the benefits and the risks of self-disclosure. According to this model, people adjust the intimacy of the information disclosed according to media affordances. When people cannot control a disclosure target (i.e., the audience), they compensate for the lack of target control by increasing information control, specifically, by sharing less intimate or private information.

## 3 UNDERSTANDING PHOTO-SHARING BEHAVIOR

We conducted an interview study to understand visually impaired people's photo-sharing practices and to inform the design of the enhanced photo-sharing feature.

### 3.1 Method

We recruited 12 participants with visual impairments (9 females and 3 males), whose ages ranged from 20 to 64 (mean=35.25), as shown in Table 1. All participants were legally blind, meaning that either (1) their best-corrected visual acuity in their better eye was 20/200 or worse, or (2) their visual field was 20 degrees or narrower [58]. Two participants (Marie, Susan) had some functional vision and used magnification on their smartphones, while the others used screen readers. All participants had experience with posting photos on Facebook.

We interviewed each participant in person in a lab environment. We asked about their experience with sharing photos on Facebook, including how often they share photos, what photos they shared, the challenges they had, and the strategies they used to overcome those challenges. We concluded the interview by asking participants what kind of information they wanted to know when choosing photos to share. Each interview lasted 30 minutes. We compensated each participant with a $50 Amazon gift card.

We transcribed and coded the interviews using Burnard's method [19]. Two researchers coded two sample transcripts separately and discussed the categories together. One researcher then coded the remaining transcripts based on the agreed categories.

Table 1. Demographic information of the 12 participants in the interview study. Participants with a "*" took part in both the interview study and the diary study. All names are pseudonyms.

| Name | Age /Sex | Visual Condition | Frequency of Photo-Sharing | The Kinds of Photos Shared on Facebook |
|---|---|---|---|---|
| Anne* | 39/F | Blind. She was blind since 13 months old. | Rare | Photos of her son |
| Helen | 36/F | Blind. She had retinitis pigmenttosa and lost her vision ten years ago. | Sometimes | Travel photos |
| Ella | 28/F | Ultra low vision. She had retinitis pigmentosa. She could see shapes, colors, and large print. | Sometimes | Travel photos |
| Gabi | 40/F | Blind. She was blind since she was three. | Sometimes | Photos of her dog |
| Jane | 64/F | Blind. She had vision as a child and lost her vision ten years ago. | Rare | Photos of herself |
| Salina | 22/F | Ultra low vision. She had Leber's Congenital Amaurosis. She had color vision and barely used her vision. | Often | All kinds of photos, such as family, travel, school activities |
| John | 20/F | Blind. He had anophthalmia. He had empty eye sockets when he was born. He had two prosthetics. | Sometimes | Photos of himself |
| Kate* | 32/F | Ultra low vision. She had a little vision in the right eye and could see shadows and light. | Once a month on average | Photos of her hobby: ceramic art |
| Peter* | 38/M | Ultra low vision. He was blind in the left eye, but had a little vision in the right eye. | Sometimes | Family photos, travel photos |
| Matt* | 32/M | Blind. He was blind since two years ago. | Sometimes | All kinds of photos, such as family, travel, selfies, food |
| Marie* | 33/F | Low vision. She could not see stuff far way or small details. She could not read small print. | Often | All kinds of photos, such as family photos, pet photos |
| Susan* | 39/F | Low vision. She had Coloboma, retinal detachment, tunnel vision, and blurry and dark spots in the central vision. | Sometimes | Photos of her dogs |

## 3.2 Findings

*3.2.1 Motivations For Sharing Photos.* All participants shared photos on Facebook. Anne and Jane rarely posted photos, but all other participants shared photos on Facebook on a regular basis. Salina and Marie showed a strong interest in sharing photos and said they uploaded photos all the time.

Similar to sighted people, participants shared various kinds of photos, including photos of family (e.g., Anne, Salina), pets (e.g., Gabi, Susan), travel experiences (e.g., Helen, Ella), and hobbies (e.g., Kate). We asked participants why they shared these photos, and identified three main reasons:

*Sharing Experiences.* Some participants shared photos to record experiences, such as family photos and travel photos. Helen described: "We went to the Golden Gate Bridge and we used Facebook to post photos and tag each other."

*Pleasing Their Audience.* Some participants shared photos to amuse their friends, caring about the audience's response. For example, when Gabi uploaded photos of her dog, she tried to find the most amusing photos. "It wasn't really the point for me to post some boring thing. I try to pick ones that people would find amusing, not just [the dog] eating, but things that sort of show him in a fun moment, that people would either laugh at or enjoy" (Gabi).

*Constructing Identities.* Some participants shared photos on Facebook to emphasize certain aspects of their identity. Kate posted photos of her ceramic art to convey her identity as an artist: "I'm a ceramic artist. I have a ceramic album on my page. I usually add photos to my collections." As another example, Peter used his Facebook profile mainly to look for jobs and communicate work-related issues, but he also occasionally uploaded family photos to show the affable part of his identity. He said, "Facebook is a professional thing to me. But I do post family photos, my wife and kid, because these are benign stuff."

*3.2.2 Risks in Posting Photos.* Not being able to examine photos posed potential risks to our participants when sharing photos on Facebook. Some participants were worried about uploading an inappropriate photo by accident. As Ella indicated, "For example, I want to post a beach picture. Please don't let me post one where like I have a wet t-shirt on." Six of the participants were concerned about embarrassing themselves and emphasizing their disabilities if they upload a "bad" or "wrong" photo. For example, Jane was embarrassed by one incident and decided not to post photos any more. "One time I did it and then somebody sent me a message and told me the photo I put was sideways. I was very embarrassed. I preferred not uploading any photos" (Jane).

Participants who used Facebook for work-related activities were concerned about posting unprofessional photos, which may negatively affect their job opportunities. For example, Peter taught adaptive technology part-time. He used Facebook to communicate with his students and look for new positions. He was careful about the content posted on his Facebook page and was worried that some inappropriate posts would harm his professional image: "I belong to several engineering and technical groups on Facebook. This is professional. This is something I do for work. So I've locked down my [Facebook] timeline pretty hard."

The potentially large audience on Facebook increased participants' concerns about potential risks. They had a strong awareness of the audience on Facebook and were careful about uploading photos for "public" consumption. As Peter said, "I'm always careful whatever I put on Facebook. Once you put it out there for the consumption of the public domain, everyone will see, and more than likely you cannot take it back."

Interestingly, compared with participants who had little to no vision, low vision participants were less concerned about sharing "bad" photos. We had two participants with low vision, Marie and Susan. Both of them did not care much about photo details (e.g., blurriness). Marie explained that she "[knows] some pictures are a little blurry, but [she] just posts them and not worry too much about the specifics."

*3.2.3 Strategies of Information Control.* Given the high risk of posting photos on Facebook, participants adopted strategies to examine and control the photos they disclosed.

Asking family or friends for help was the most common strategy participants used. They usually asked sighted people to describe or verify the photos in person before uploading to Facebook. However, many participants were burdened by having to ask for help and wanted to be more independent. "Having a sighted friend [help me], I just don't want to do that all the time. It gets kind of annoying for me. I don't mind asking for help but sometimes I want to do it on my own" (Kate).

Some participants (e.g., Salina, Peter) used TapTapSee, a smartphone application that combines computer vision and crowdsourcing to describe photos for blind people [59]. However, the quality of the descriptions from TapTapSee could not be guaranteed. The descriptions sometimes were too general to be useful for a blind user. Salina complained about the inconsistent quality of the descriptions, "Sometimes it's good, and sometimes it's not. Once it said, 'white packet with blue letters.' Why don't you tell me what those letters say since you just read them?"

Some participants avoided uploading photos to make sure they did not disclose anything inappropriate. For example, Anne carefully examined the contents posted on her Facebook timeline: "Unless I am really sure what it is, I tend not to upload photos. It makes me uncomfortable."

*3.2.4 Information Needs.* We categorized the kinds of information that participants wanted to know about photos before sharing them:

*Key Visual Elements.* All participants wanted to know the key visual elements in a photo, both in the foreground and the background. As Helen described, "It would be cool if I was taking a picture and the Golden Gate Bridge was there, it would say, 'the Golden Gate Bridge in your picture,' or 'clouds in the sky," or 'a train in the background,' or some of those key elements."

*People.* Nine participants wanted to know detailed information about people in the photo, especially their identities. Anne explained: "If I want to upload [a photo of] my son but it was my mom's picture, it would be embarrassing." Anne and Ella were also interested in people's relative location in the photo.

*Photo Quality.* Participants cared about the quality of their photos. Five of them suggested adding quality-related information to the photo descriptions, such as the quality of the lighting and the degree of blurriness. For portraits, they wanted to know the quality of the faces in a photo, for example, whether someone's face is cut off, people's eyes are open, or their facial expressions are appropriate. "I always need to confirm if everybody's eyes are open. I want to have a picture where people are smiling obviously, a picture of the best lighting, and also a picture that is crisp" (Ella).

## 3.3 Discussion

Visually impaired people wanted to share photos on SNSs and they did so for the same reasons that sighted people share photos [30,32]. Much prior work has discussed people's motivations and behaviors for photo sharing in the context of self-disclosure and self-presentation theories [10,21,44]. All existing work focused on sighted people. We analyze the photo-sharing behaviors of visually impaired people, contributing to these theories from a new perspective.

Self-disclosure and self-presentation are closely related. In the photo-sharing context, people are disclosing their personal information via photos and presenting a desirable image to the public at the same time. When disclosing photos on SNSs, people benefit from the rewards of self-presentation, but also face risks from self-disclosure.

Prior work discusses how sighted people share photos on SNSs for self-presentation. Similarly, we found that although visually impaired people do not "see" their and others' photos, they still constructed their identities and developed relationships with their audience through this visual medium. Specifically, we identified three main reasons that motivated our visually impaired participants to share photos: sharing experiences, pleasing their audience, and constructing identities, which aligned with the goals of self-presentation [8]. This indicated that visually impaired people were eager to present themselves through photo sharing and reaped the same rewards from this form of self-presentation as their sighted peers.

Beyond self-presentation, photo sharing also involves self-disclosure, which poses certain risks. Compared to sighted people, visually impaired people faced greater risks when sharing photos on SNSs. Prior research [10] identified vulnerability and information loss as potential risks for sighted people. For example, disclosed information can lead to unfavorable impressions of the discloser, and may even lead to a shift in the power dynamics in the relationship between the discloser and the recipient. Our study showed that such risks also applied to visually impaired users when sharing photos online, and were even more severe due to their disabilities. Our participants were particularly concerned about sharing inappropriate, unintended, or low quality photos because this would emphasize a disability-related vulnerability. Many people with disabilities do not identify themselves as 'dis-abled' [20,48]. While accepting their physical impairments, they consider themselves as capable as other people and wanted others to perceive them as such. Thus, sharing uninteded photos posed a high risk of presenting their disability as a barrier in a daily activity, which could in turn impact important aspects of their lives, such as their relationships with others, and even job opportunities. The massive and loosely defined audience on SNSs exacerbated this risk.

Our study highlighted a potential difference between the way blind and low vision people perceived the risk associated with posting low quality photos. Compared with blind participants, participants with low vision cared less about posting "blurry" photos. This may be because low vision is often an invisible disability, meaning that people do not usually appear "disabled" to others [43]. As a result, they are less concerned with presenting a capable image on SNSs. More work is needed to investigate this preliminary finding, however, since we only had two low vision participants in our study.

With the risk of exposing disability-related vulnerabilities and limited access to the visual content of photos, visually impaired users face greater challenges than sighted people in curating different aspects of their self-identity online. On one hand, they need to identify the photos that can help construct their preferred identity (e.g., ceramic art photos for Kate's identity as a ceramic artist). On the other hand, they need to filter out the low-quality photos that may highlight disability-related difficulties.

To address these challenges, visually impaired people sought assistance from family, friends, or human-powered services to examine photos before posting. This aligns with self-disclosure theory which states that people may compensate for the lack of audience control (the large audience on Facebook) by increasing information control [10]. "Information control" involves controlling the intimate or private information disclosed by the shared content [10,11]. In our case, however, visually impaired people wanted to control information even when it was not intimate and not private because they could not see the photo. Thus for visually impaired people, information control also involves rigorous evaluation of photo content and quality. However, the only method for visually impaired people to conduct information control was to seek sighted assistance, which contradicts the capable and independent impression they seek to construct, causing a psychological toll and social costs [18,22,38]. This revealed a major gap in SNS accessibility.

### 3.4 Design Implications

Visually impaired people want to construct an impression of competence by sharing specific photos online. However, they have to rely on human assistance to complete this task, which is contrary to the capable image they want to construct. Computer vision technology has the potential to ease this tension by providing access to visual information to help people select photos independently, or at least reduce the amount of work required from human assistance.

According to the interview study, a good photo description needs to include three categories of information to enable visually impaired users to decide which photos to share: key visual elements, people, and photo quality. While the need for information about key visual elements and people is consistent with findings reported by Wu [51] in the photo consumption use case, the importance of photo quality information is much more pronounced for photo sharing. As mentioned by our participants, these aspects could include, whether the photo is blurry, whether it has good lighting, or whether people in the photo are smiling with open eyes.

## 4 ENHANCING THE PHOTO-SHARING FEATURE

Guided by the design implications from the interview study, we designed and built an accessibility feature in the photo-sharing interface in the Facebook Android application, which automatically generated photo descriptions in a user's local album and verbally reported the descriptions to visually impaired users via a screen reader. Our goal was to enable visually impaired people to browse their albums and share photos on Facebook more independently.

### 4.1 Photo Description Design

We designed the photo descriptions to support participants' needs based on what we learned in the interview study. Our descriptions included three components: key visual elements, people, and photo quality. In the current prototype, the photo-quality information is only related to human faces.

To describe the key elements in a photo, we used the descriptions generated by automatic alt-text (AAT) [51]. AAT recognizes 97 visual concepts, describing an image with four categories of concepts, including people (e.g., people count, child, baby), objects (e.g., car, building, tree, food), settings (e.g., inside restaurant, outdoor, nature), and other image properties (e.g., close-up, selfie, drawing). These concepts were selected by three criteria: (1) prominence: whether a concept is the salient part of a photo, (2) clarity: whether a concept is easily and clearly defined without disputations, and (3) coverage: whether these concepts can cover all categories of the majority of photos. Similar to alt-text [51], the precision of each concept was manually evaluated with randomly sampled public photos on Facebook. All concept tags reached a precision of 0.8 or above (a few more sensitive concepts have precisions as high as 0.98) with various recalls. Over 75% of the randomly sampled photos receive at least one concept tag. Note that the photos used to train and test the algorithm were shared photos on Facebook, which were relatively high quality compared with photos taken by people with visuall impairments. In the future, we will train and evaluate our algorithms on photos taken by visually impaired people to achieve higher accuracies for our use case. Fig. 1a and 1b give two examples of computer-generated descriptions for key elements in photos taken outdoors and indoors.

Information about people was also important for visually impaired users when choosing photos to upload. To support this need, we applied state-of-the-art computer vision technology to detect faces and ran face recognition[2] to identify people in photos during this research study. To protect the privacy of people in the photos, we only recognized a person if he was one of the user's Facebook friends and had the "tag suggestion" feature[3] enabled, otherwise, we referred to him as "unknown person." Our face recognition model was designed and trained in a similar way to the one described in Zhang's work [52] and reaches an accuracy of 97% on a public Flickr photo dataset.

We added photo quality information to help visually impaired people select photos. In the current prototype, we focused on providing quality-related information for human faces in a photo because faces are usually the most important part in a photo that includes them. For the faces detected in a photo, we ran several image classifiers to detect different characteristics of each face, including whether the face was blurry, whether the person was smiling, and whether his or her eyes were open, as shown in Fig. 1c. Those classifiers were trained separately and reached a precision of 0.9 or above.

Similar to AAT, we organized the recognized information into a complete sentence, as it sounds more natural and "friendly" [51]. The sentence always starts with "Image may contain" and is followed by information about people, objects, and the general setting or properties of the image. We used the following structure to present detailed information about people and facial characteristics: people count, [person1: identity, other attributes], [person2: identity, other attributes]. The people in a photo are described in order from left to right. For example, in Fig. 1c, there are two people recognized in the photo and the generated description would be "*Image may contain: two people, name1, not blurry, smiling, eyes open [person1 on the left], name2, not blurry, smiling, eyes open [person2 on the right], eye glasses, text, indoor.*"

## 4.2 Interaction Design

The description feature works with TalkBack [27], the built-in screen reader on Android that allows visually impaired people to use touch gestures (e.g., tap and swipe) to get audio feedback about the content of the phone user interface. Android applications can set the alternative text field of each virtual component on screen to make it accessible with TalkBack. When a user sets the focus on a component, TalkBack reads the alternative text for that component aloud. A user can tap on the component or navigate to it with swipe gestures to set the focus. We designed the interaction for the description feature based on TalkBack interaction to shorten the learning curve.

In the current prototype, the photo description feature is only available in the Facebook composer feature when sharing photos (Fig. 2a). When a user wants to post a photo, she can load her photo gallery via Facebook and access the descriptions. We used the computer-generated descriptions as alternative text for each photo. Originally, if a user taps on a photo, TalkBack only reads "image, in list." With our prototype, when a visually impaired user sets the focus to a local photo, the Facebook application automatically sends the photo to a remote server for image recognition and generates a photo description, which is read back by TalkBack to the visually impaired user (Fig. 2b). The photo is immediately deleted from the remote server once the description is generated. While the photo is recognized, the composer verbally announces "generating description" to indicate that the user should wait for the recognition results. The gestures to set the focus are the same as with TalkBack: tap on a photo, or navigate to a photo with a swipe. After the first recognition, the system caches the description locally so that the user can hear the description immediately the next time she focuses on the photo.

After a user selects a photo with a double-click gesture (the TalkBack gesture), she can still listen to the description by setting the focus to this photo. The system will also inform her that this photo is selected by adding "selected photo" before the original description (Fig. 2c). The user can post all the selected photos by clicking the "Done" button on the upper right corner (Fig. 2d).

---

[2] Note that face recognition was only used for this non-commercial research project. We did not use face recognition in AAT on Facebook.
[3] When a friend uploads your photo, Facebook may recognize you and suggest tagging you in the photo. You can choose whether or not Facebook suggests your name by turning on or off the 'tag suggestion' function.

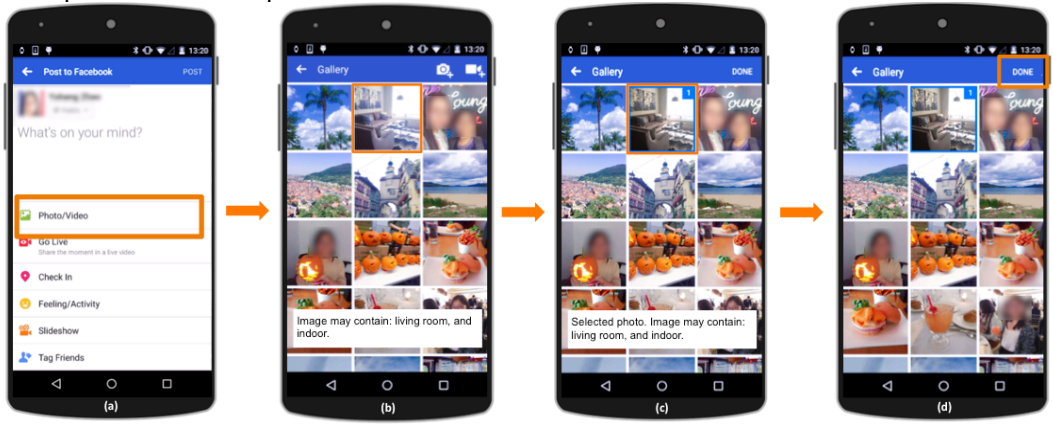

Fig. 2. The workflow of the photo-sharing feature: (a) The entry point to the photo-sharing feature in Facebook; (b) A user can get a photo's description by setting the focus on a photo; (c) After a user selects a photo to upload, by tapping this photo, she can still get the photo description; (d) When completing the photo selection, a user can double click the "Done" button to post the photos.

## 5 EVALUATION

We evaluated the enhanced photo-sharing feature with a seven-day diary study. Our goal was to evaluate the effectiveness of the computer-generated descriptions and understand their effect on the photo-sharing experience of visually impaired people.

### 5.1 Method

*5.1.1 Participants.* Six participants (2 males and 4 females) from the interview study volunteered for the diary study. Table 1 shows their demographic information (the participants with a "*" took part in the diary study). Participants' ages ranged from 32 to 39 (mean=36). They were all legally blind. Two participants (Marie, Susan) had enough functional vision to use magnification on their smartphones, while the others used screen readers. All participants had experience with Android phones. They are all Facebook users and have posted photos to Facebook before. The diary study lasted seven days. Participants were compensated with a $100 Amazon gift card for each day of the study.

*5.1.2 Procedure.* The study consisted of three parts: a training session, a diary study, and an interview.

The training session lasted one hour in a lab environment. We asked about participants' demographic information and then demonstrated the new feature. If the participant had an Android phone, we installed the experiment version of the Facebook application with the description feature on their phone. If not, we gave them a new Android smartphone (a Nexus 5) with the experimental Facebook application with the enhanced feature. We asked the participants to practice using the computer-generated descriptions until they could confidently access photo descriptions and share photos on Facebook. One participant (Susan) had not used TalkBack before, so we trained her to use TalkBack and made sure she memorized the gestures to access the photo descriptions.

Participants used the enhanced photo-sharing feature for one week. We instructed them to use it for at least four days during the study. We sent them an eight-question survey (Table 2) each day, asking about their experience with the computer-generated descriptions for that day. They were asked to fill out the survey every day, including the days in which they did not use the descriptions. We instrumented the experiment application to log when participants used the descriptions and how many times they uploaded photos. Participants were advised to email or call us at any time during the study for troubleshooting.

Table 2. The eight-question daily survey for the diary study

| | |
|---|---|
| Q1 | Did you use the description feature today? |
| Q2 | Did you post a photo for Facebook using the computer-generated descriptions? |
| Q3 | How many photos did you examine before deciding which one to post (1, 2-5, 6-10, 11 or more photos)? |
| Q4 | What type of photos did you upload to Facebook? |
| Q5 | How useful were the descriptions today (extremely, very, moderately, a little, not at all)? |
| Q6 | What type of information provided by the descriptions was most useful for you to decide which photos to post? |
| Q7 | What can be done to improve the descriptions? |
| Q8 | If you did not use descriptions, why didn't you use it today? |

Lastly, we conducted a follow-up interview to collect participants' feedback on the description feature, including how effective and accurate the descriptions were and how they affected participants' photo-sharing experiences. Participants were asked to give scores to the effectiveness and accuracy of the descriptions, ranging from 1 to 10 (1 represents not effective/accurate at all, while 10 represents extremely effective/accurate). We also asked them how the descriptions could be improved.

We concluded the interview by observing how the participants used the description feature on their own to understand their experiences.

*5.1.3 Analysis.* We video recorded the training and interview sessions. The videos were transcribed by a professional service. We coded the transcripts using Burnard's method [19]. Two researchers coded two sample transcripts separately and discussed the themes and categories together. One researcher coded the remaining transcripts based on the agreed categories. We also aggregated and coded the participants' answers from the diary surveys to understand their daily experiences.

We computed the number of times that participants browsed and uploaded photos during the week based on the logged data. Since browsing photos is a continuous process, we treated the logged data in a continuous time period as one browsing session. Browsing sessions needed to be at least 30 minutes apart to be counted as distinct sessions.

## 5.2 Results

*5.2.1 Use of the Description Feature.* On average, participants used the computer-generated descriptions to browse photos in their albums six times but only posted photos on Facebook twice throughout the study. Fig. 3 showed the number of times that each participant browsed and uploaded photos during the whole week. We found that the number of times participants posted photos was much smaller than the number of browsing sessions. Matt especially browsed photos more times than all the other participants, but never shared one photo to Facebook. We asked him about what prevented him from sharing photos and found that he just did not have any photos he wanted to share during that week. As Matt explained, "I just didn't have anything to post. I don't do that like, very often." This revealed one of the potential benefits of the photo descriptions: even when participants did not have photos to share, they still wanted to use the descriptions to explore their photos and recall memories. We explain this more in the "Reappropriation" section.

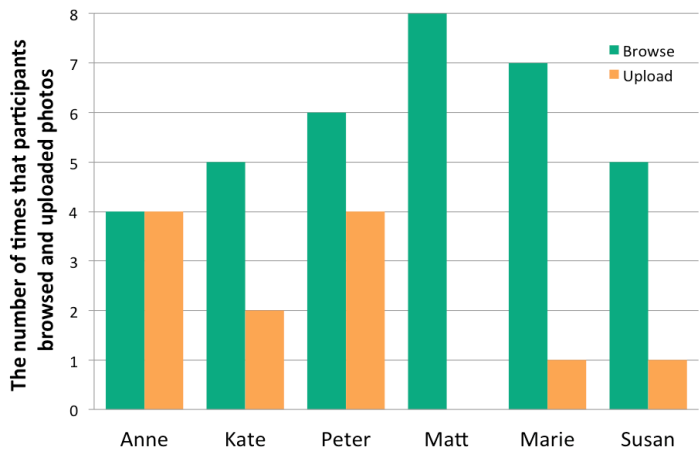

Fig. 3. The time that each participant spent browsing (green) and uploading photos (orange) during the week.

According to the self-reported survey data, participants usually examined 2-5 photos before finding one to upload. However, Marie once examined 6-10 photos, and Anne sometimes only examined one photo before uploading, because she often took a photo and immediately uploaded it. She used the descriptions only to confirm the content of the most recent photo.

Participants shared different kinds of photos during the diary study. Four people posted photos of animals, three posted photos of objects (e.g., ceramic flowers, food, paintings), two posted photos of people, and one posted a photo of a landscape (a baseball stadium).

*5.2.2 Effectiveness.* All participants agreed that the computer-generated descriptions were useful in their photo-sharing experience. However, the scores they gave to the effectiveness of the descriptions in the follow-up interview varied, ranging from 3 to 10 (mean=7, SD=2.61). Most participants gave relatively high scores (7-10), while two gave low scores (Anne: 5, Susan: 3).

Most participants were excited about the computer-generated descriptions and thought they fulfilled their needs by providing information about the key elements and people in a photo. Peter gave an example of a photo he took at a baseball stadium, "It was really neat because it would tell me where I was, what it looked like, and picked out objects. It told me 'Stadium, outdoor, and baseball.' It's perfect." Matt expressed his excitement about face recognition, "I always hoped that facial recognition could be built into the alt-text set up. This is what I was looking for." Participants also benefited from the descriptions of facial characteristics. As Marie said, "I liked it because sometimes I can't tell if a person's eyes are closed. So that was kind of cool to know, my eyes are closed. So that's not a good photo.'"

However, Anne and Susan felt the descriptions were not rich enough to help them share photos independently. As Anne explained, "It gives me a little more idea than before, but it wasn't helpful enough that I can do it on my own."

Participants were asked about the most useful information provided by the descriptions in the daily survey (multiple selections were allowed). Fig. 4 shows the number of times that each type of information was selected. Information about face blurriness was selected more than other categories of information, highlighting its importance for photo sharing. Key visual elements were also selected often (6 times).

When we investigated why face blurriness was selected most often, we found that it was especially important for people who still had a little functional vision (e.g., Marie, Susan). When navigating an album, they were able to perceive the general content of the photos, but struggled to determine whether the photo was blurry. As Marie described, "Sometimes photos are just slightly blurry and I won't know. So details like blurriness are more useful." This is interesting because both of our low vision participants mentioned that they did not care about sharing photos that were a little blurry in the first study. Besides faces, participants also wanted to extend the blurriness feature to the general photo.

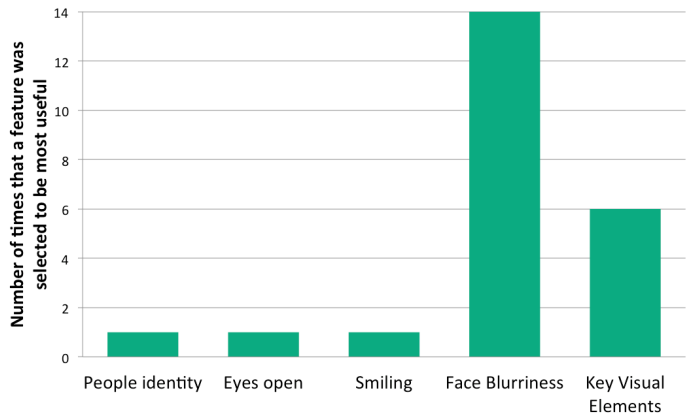

Fig. 4. **The number of times that each type of information was selected to be most useful in the daily survey.**

*5.2.3 Accuracy.* All participants agreed that the accuracy of the computer-generated descriptions affected their experience most and that "it really drives the helpfulness" (Kate). While the accuracy of our algorithms was high (described in "Description Design"), participants' perceived accuracy (i.e., how accurate people felt the descriptions were), varied, ranging from 0.1 to 0.9 (mean=0.63, SD=0.29). Four people estimated a relatively high accuracy (0.7-0.9), while the other two reported low perceived accuracy (Anne 0.5, Susan 0.1). In fact, Anne and Susan were the only two people who gave low scores to the descriptions' effectiveness, indicating that the perceived accuracy likely impacted the usefulness of the computer-generated photo descriptions.

In addition to the perceived accuracy, we also asked participants what level of accuracy they were willing to tolerate to feel comfortable using the photo descriptions for photo sharing—the acceptable accuracy [35]. The four participants who gave a relatively high perceived accuracy all felt the current accuracy was acceptable. Anne felt an accuracy of 0.8 was acceptable, while Susan was not sure about the acceptable accuracy. In general, participants believed that the accuracy around 0.8 would be acceptable for an effective description in the photo-sharing context.

Most participants felt that the application was fairly accurate and were amazed by the computer-generated descriptions. "It was amazing. It was like 'outdoor, person, smiling, eyes open,' it would tell me all the things and it was right. It also said, 'not blurry'" (Marie). However, Anne and Susan were not satisfied with the recognition accuracy. Anne felt confused after receiving a wrong recognition, "Am I choosing the right photo? Was this from something else that I took before? I had to ask my husband to confirm before I posted it. I don't trust it."

We found that photo-quality issues led to a low recognition accuracy. Since taking photos is a major challenge for visually impaired people [31,32,45], some photos in a participant's album had quality issues, such as blurriness or low luminance. Some participants noticed the relationship between recognition accuracy and photo quality during the study. Peter, who had a technical background mentioned, "I noticed that the accuracy depended on the lighting. It seemed to be the biggest factor affecting the accuracy."

The design of the descriptions for photos with quality issues was another reason that caused the participants to perceive a low accuracy. For example, Susan conducted recognition on blurry photos of moving people and was disappointed when they were not recognized: "There were some pictures that were blurry because they were kind of action shots, but you could still tell that they were people." In fact, our algorithms could recognize the shapes of people, but the blurriness reduced the confidence score of the recognition result, so it did not pass the precision threshold. This suggested that we should adjust the descriptions based on the photo condition, for example, informing the user that the photo has quality issues instead of simply dropping uncertain recognition results.

Participants also suggested providing certainty indicators of the information in the descriptions to handle inaccurate recognition. As Susan said, "If you are sure, you can say, 'this photo probably contains a dog.' If not, it can be 'this photo might contain a dog.'" This echoed the result from MacLeod's study [37], which emphasized the importance of communicating the uncertainty of computer-generated descriptions to visually impaired people.

*5.2.4 Descriptiveness.* Descriptiveness was another factor that affected participants' experiences with the descriptions. We report their opinions on the descriptiveness in terms of the level of detail and the salience of photos.

Level of detail. Participants thought some of the descriptions were too general and wanted more descriptive details about salient entities in the photo. Some participants wanted to know the characteristics of an object (e.g., kind, color). For example, Kate took photos of different kinds of flowers with different colors but the descriptions were only "flower." She wanted the descriptions to provide more details about the color and the kinds of flowers. Besides details about objects, participants were also interested in the behaviors of people or animals in the photo. As Susan suggested, "All it says for the dog is 'dog.' I want to see which direction is the dog looking, and what they do, is their tongue out or something." The need for details was especially strong for people with low vision (e.g., Marie, Susan). "As a low vision person rather than completely blind, I knew I was taking a picture of the dog. So for me, the general description isn't that useful" (Susan).

Since the descriptions were too general, participants had difficulty differentiating among similar photos because often they all had identical descriptions. They wanted the descriptions to help them decide which photo to post. For example, Peter had several similar photos of his room, which were all labeled with "indoor." He said: "It's just saying indoor. It's not telling me anything else. At least try to tell me some differences, so I know which is which."

Salience. Participants cared about certain visual elements that were not necessarily salient when removed from the context of the SNS. For example, Anne had a photo of herself standing and holding a small plate of fruit in her hand, and the composer only described the person in the photo. She was not satisfied with the description because it did not include the fruit. She explained: "One person, [name], no blurry, no smiling, eyes open, indoor. It's correct. But see, this one doesn't tell me I have some fruit in my hand and what fruit it is. So I wouldn't know the differences between this photo and the other photos [of me]." In the context of Anne posting photos of herself, it was more important to communicate the fact that she was holding something rather than that she was in her photo.

*5.2.5 Personality.* When deciding which photos to share on Facebook, participants cared about whether and how their personalities were presented in a photo and wanted the descriptions to provide relevant information.

Personal Style. Participants wanted to share photos that can present their individual styles. For example, Susan wanted to upload more interesting photos of people instead of a traditionally "good" photo with a big smile. As she indicated, "I know whether people are smiling or not [from the descriptions], but that's a little bit harder to determine whether I would want to share because sometimes you want people to look goofy or whatever."

Personal Aesthetic. Beyond objective descriptions, participants wanted to know whether photos fit their personal aesthetic. "It's very subjective on what's a good picture and what's not because people have different artistic views. It might be hard to totally rely on [the computer-generated descriptions]" (Anne).

*5.2.6 Trusting the Descriptions.* Participants did not completely trust the computer-generated descriptions so they used different means to verify their correctness. Most participants asked sighted people to double check whether the descriptions were accurate. Some participants (Kate, Peter, Matt) memorized the general content of their photos and verified the descriptions by checking whether they matched their expectations. "As long as the description was in the way that I thought the picture was supposed to be, I felt confident about it" (Matt). Participants with relatively better vision (Marie and Susan) also magnified the photo and checked it by themselves.

Despite these strategies, asking for sighted assistance was the only method that most participants relied on when sharing photos on Facebook. This partially explains the findings described in the previous section that participants used the description feature many times to browse photos, but rarely uploaded any photos (Fig. 3; although another reason was that some participants simply did not have any photos to share, but just wanted to explore their photos). When asked whether they would post a photo directly if they felt the description was correct, all participants indicated that they still preferred a sighted person to confirm.

We found that participants had a strong awareness of the large audience on Facebook and the potential risks of sharing photos to the public. As a result, the demand for good descriptions in terms of accuracy and descriptiveness was high; participants were reluctant to trust the current descriptions. Anne explained:

*"One hesitation around using the tool is that, it's a high bar to clear before you post photos, especially on Facebook. There are many hundreds of people out there. I don't want to make a fool out of myself. You have to have a lot of faith in the tool before you want to potentially take that risk."*

*5.2.7 Auxiliary Information.* Besides the descriptions of the photo content, participants wanted to know more auxiliary information to help them recall memories and decide which photos to upload.

Photo Quality. Participants expected more information about the photo quality. The level of luminance of the photo was important to them. As Matt indicated, "I want to know the luminance of the photo because I've definitely taken some photos where I thought my light was on but it's actually super dark." Moreover, Susan wanted to know the specific level of blurriness because she thought a little blurriness is acceptable. "I'm not totally opposed to posting something that's a little blurry. But they don't have a degree. They just say blurry or not blurry."

Time and Location. Some participants suggested using timestamp and location information to help them navigate their albums. After getting many photo descriptions, it was difficult for the participants to locate their current position in the album. They hoped to get the timestamps of each photo to better understand which photo they are investigating. As Matt suggested, "You can add timestamps, so that I can know how the photos are sorted and which photo I'm tapping." Peter also wanted to extract location information from the GPS to help him recall memories related to certain photos.

*5.2.8 Reappropriation.* Although participants did not want to upload photos using the computer-generated descriptions without sighted assistance, they found the descriptions very useful in other ways: recalling memories and organizing photos.

Some participants used the descriptions to recall memories. When they took a lot of photos at one time and did not have time to label each one, they used the computer-generated descriptions to revisit the photos and reminisce. As Matt said, "I have a bunch of photos and I don't necessarily remember what I was taking a photo of. Using the descriptions as a tool to look through my gallery is extremely interesting to me."

Participants also filtered and organized their local photos by leveraging the computer-generated descriptions. They could delete imperfect photos and clean their albums based on the descriptions. As Marie indicated, "I want to delete a lot of photos because my storage will get full. If the description told me a photo is blurry, I can delete it. That was helpful."

While the current description feature was only provided in the Facebook composer, participants wanted to extend it to a smartphone feature that could be applied to their local albums. "[I hope the descriptions] would be able to hop into my actual gallery. That would be perfect" (Matt).

## 6 DISCUSSION

We evaluated the effectiveness of the enhanced photo-sharing feature and found that the computer-generated descriptions were useful for recalling memories and organizing photos for visually impaired people. However, when sharing photos on SNSs, people did not trust the descriptions and still heavily relied on sighted assistance.

### 6.1 Descriptions, Photo-Sharing, and Self-Disclosure

As discussed in the interview study, visually impaired people faced additional risks compared to sighted people when disclosing photos on SNSs. Posting an unintended or inappropriate photo may emphasize a disability-related difficulty and present an image of incompetence to a large audience. Thus, they employed an information control mechanism—asking for sighted assistance—to compensate for these high risks. In our study, we hoped that the computer-generated descriptions would act as a new information control strategy, replacing the use of sighted assistants. However, different information control strategies accrue corresponding costs. We consider the effectiveness of the computer-generated descriptions by comparing its reward-cost ratio with that of sighted assistance.

To optimize the reward-cost ratio, people choose the strategy with lower costs given that the rewards of self-disclosure on SNSs are the same. In the photo-sharing context, the cost of using sighted assistance includes the social costs of asking for help from a specific person [18,22], while the cost of using computer-generated descriptions was the risks of revealing the visually impaired user's difficulties to the public by posting a low quality or unintended photo if they trust an inaccurate description. According to our findings, the latter cost was much higher than the former since they were afraid of "making a fool of themselves." This yielded extremely high demands on the computer-generated descriptions in terms of both accuracy and descriptiveness, which explains why participants had difficulty trusting the descriptions.

Self-disclosure theory can also explain why computer-generated descriptions have a different effect on photo sharing (content production) and photo consumption. AAT [51] used computer-generated descriptions to help visually impaired people perceive shared photos on Facebook. Although trust issues arose because of occasional recognition inaccuracy, participants enjoyed the descriptions and self-reported being more likely to "like" and "comment" on photos when using AAT. This is because the risk of mis-reacting to other people's posts is much lower than that of mis-disclosing one's own information on SNSs.

## 6.2 Collaborative Photo Sharing

The computer-generated description feature was designed to enable people with visual impairments to share photos on SNSs more independently. However, given the perceived risks in uploading photos to sites with potentially large networks, there is a reluctance to rely exclusively on the descriptions, especially when there are any doubts about its accuracy. Our study showed that all the participants needed to confirm photo contents with sighted people before uploading photos.

Although they did not upload photos without sighted confirmation, our participants used the description feature to get information about their photos and organize their albums. Our study showed that the participants enjoyed using the descriptions to understand photo contents, recall memories, and delete photos with low quality (e.g., photos with blurry faces), highlighting the capability of computer-generated descriptions to increase photo-sharing efficiency and reduce workload for sighted assistance.

The potential of computer vision technology and users' trust of sighted people suggest an opportunity for collaborative photo selection, where computer-generated descriptions and trusted sighted examination is combined. One limitation of previous attempts to facilitate photo sharing, such as VizWiz Social [18], was that people were concerned about creating an undue burden on their friends or not appearing independent. A sharing tool like the one discussed in the present paper could help alleviate these concerns, especially if it was paired with existing SNSs. For example, a person with a visual impairment could use the computer-generated descriptions to narrow down a large set of photos, eliminating ones with obvious issues (e.g., blurry photos). Once the photo set has been narrowed, she could then consult a friend in her network to select the best photos to upload and share more broadly.

Such collaborative photo selection would both allow people with visual impairments to feel more independent and engaged in photo sharing, while also reducing concerns about burdening friends to sift through large sets of photos.

## 6.3 New Directions For Computer Vision Research

Our description feature used state-of-the-art computer vision technology, but the algorithms were the same as those used in alt-text [51], which were designed and trained to recognized the content of good-quality photos that are already shared on SNSs. These algorithms may not be suitable for a photo-sharing context, where users may have many similar photos and the photo quality could not be guaranteed. Algorithms on photo quality detection and photo differentiation should be used to increase the usability of photo descriptions. We should also train the object recognition algorithm with photos taken by visually impaired users to improve its precision for this use case.

Conventional approaches to automatic photo description generation did not capture certain aspects of information that are needed in a photo sharing use case. Although some existing photo recognition tools, such as Google Photos [60] provide functions to select the "best" photos and retouch photos, they were not designed for visually impaired people. They do not provide specific descriptions that explain why some photos are better than others (i.e., the specific differences among them) and how the photos were corrected. This limits visually impaired people's ability to learn about the contents and quality of their photos and control the information they post. Indeed, one of the insights from our work is that the definition of a "good" description depends heavily on the use case and the user's preferences. Our diary study spurred several new research directions for computer vision research that will yield more effective descriptions for photo sharing in an accessibility context.

*Photo Differentiation.* In our study, participants sometimes had similar photos of the same people or objects in their albums and needed to select the best one to upload. However, the recognition algorithms tended to provide identical descriptions since the photos contained similar visual elements. We propose that researchers design computer vision algorithms that could identify and describe the fine-grained differences among a group of similar photos. Such differences should include differences in quality as well as the composition of the photo, for example, one phone has flowers at the center, and the other has flowers on the left.

*Contextual Saliency Detection.* Salience detection is a popular topic in the field of computer vision, in which researchers detect visually salient entities that align with people's visual perception or attention (e.g., [1,42]). However, we found that participants' interest in particular visual elements of a photo varies depending on the context (the time of sharing, the person sharing it, other SNS activity, etc.). As a result, object saliency could often be context sensitive. For example, when a user posted a photo of herself holding a bowl of fresh fruit, the most salient object is normally the person and her face or gesture. However, if the caption of the photo says "guess what I get from the farmers' market?", or if the user is known for not liking fruit, the bowl of fruit becomes the more prominent part of the image. Contextual information could be crucial for the next generation of salience detection algorithms in compute vision.

*Personal-Aesthetic Recognition.* Although researchers have been working on recognizing artistic styles in images (e.g., [25,41]), participants in our study preferred to know whether a photo fit their personal aesthetic. Since people's aesthetic can be very subjective, we suggest building models based on a user's photo-sharing history to describe how well the photos reflect their own aesthetic style.

*Dataset Construction for Visually Impaired People.* Visually impaired people also take photos, but their photos may include different contents and have different quality from those taken by sighted people. Existing image datasets such as ImageNet [61] or COCO [62] only include photos taken by sighted photographers, which may not be suitable for image recognition for visually impaired users. Kacorri et al. [34] tried to address this problem by designing personal object recognizers that enable visually impaired people to train recognition models with a few snapshots. However, this requires users to spend time taking photos to train the models. We suggest constructing an image dataset of photos taken by visually impaired people to better support photo descriptions for visually impaired users. Although prior research constructed the VizWiz dataset [63] by collecting photos uploaded with associated questions by visually impaired people through the VizWiz application [13], the photos were organized according to the question types. A dataset organized by photo contents (e.g., object categories) and photo quality for the purpose of computer vision is needed to support description generation for visually impaired people.

*Evaluation with Human Factors.* Although the state-of-the-art image recognition algorithms we used achieved high accuracies in benchmark datasets [51], participants perceived much lower accuracies of the computer-generated descriptions. Recognition algorithms are typically evaluated with a set of stock images, that are (intentionally) not designed for a specific application. Our work has emphasized the importance of evaluating recognition algorithms in situ with specific applications.

## 6.4 Limitations

Evaluating photo-sharing behaviors is challenging. Sharing photos on an SNS only happens organically when a user has an interesting photo to post, feels the need to share the photo to a large audience, and has time to post it. We chose a diary study and an interview to capture and understand behavior patterns to the best of our ability. However, the diary study did have some limitations: it only included six participants and each participant completed an average of six browsing sessions and posted only two sets of photos. Despite these limitations, our work raised important insights and directions for future research that can improve accessibility in a time where content production is a powerful tool in different facets of life.

Our studies focused on visually impaired users and their experiences sharing photos. However, photo sharing on SNSs is an interactive experience between the sharer (in our case, the visually impaired user) and their audience. We did not explore the audience's experiences or reactions to photos directly in this paper, relying only on our participants' perceptions of their audiences. This limitation raises many interesting questions for future work involving, for example, perceptions of ability on social media. Does viewing a "bad" photo that is shared online result in a negative impression of the poster's general competence? Do people "like" or comment differently on photos shared by visually impaired people vs. sighted people? How does photo quality and contents affect this kind of engagement?

## 7 CONCLUSION

Our goal was to facilitate photo sharing for people with visual impairments. In this paper, we described how we enhanced the photo-sharing feature in the Facebook mobile application with computer-generated descriptions. We conducted an interview study to inspire our design and a seven-day diary study to evaluate the effectiveness of the description feature. We found that the descriptions helped users recall memories and organize local photos, but participants did not fully trust them when sharing photos on Facebook. They still preferred sighted assistance to share the right photo. We discussed how computer-generated descriptions affected people's experience in photo sharing through the lens of self-disclosure and self-presentation theories and also proposed new directions for computer vision research to better facilitate visual content sharing for people with visual impairments. We hope our work can inspire researchers and designers to design more effective tools to facilitate visual content sharing for people with visual impairments.

## ACKNOWLEDGMENTS

We thank our colleagues at Facebook, especially, Brett Lavalla, Jeff Wieland, Matt King, Alex Dow for their insights and guidance on the design and implementation of our prototype, and Jean Niehaus and Norberto Andrade for the helpful feedback on the paper draft. We thank the LightHouse for the Blind and Visually Impaired for the assistance on participant recruitment and providing the space for the interview studies. We also thank Mor Naaman, Serge Belongie, Baoguang Shi, Yin Cui, and Xiao Ma from Cornell Tech, who provided valuable feedback to our paper.